\documentclass[prl,twocolumn,showpacs]{revtex4}

\usepackage{graphicx,epsf}
\usepackage{bm}

\begin{document}

\title{Nonequilibrium Noise as a Probe of the Kondo Effect in
Mesoscopic Wires}
\smallskip

\author{Eran Lebanon and P. Coleman}
\affiliation{Center for Materials Theory, Serin Physics Laboratory,
    Rutgers University,\\
136 Frelinghuysen Road, Piscataway, New Jersey 08854-8019, USA}

\begin{abstract}
We study the non-equilibrium noise in mesoscopic diffusive wires
hosting magnetic impurities. We find that the shot-noise to current 
ratio develops a
peak at intermediate source-drain biases of the order of the Kondo
temperature. The enhanced impurity contribution at intermediate
biases is also manifested in the effective distribution. The
predicted peak represents increased inelastic scattering rate at
the non-equilibrium Kondo crossover.
\end{abstract}

\maketitle


Recent advances in submicron physics have opened up a new window
into the collective quantum physics of the electron gas away from
equilibrium.  One of the interesting open questions concerns the
many body physics of magnetic moments coupled simultaneously to
two Fermi gases with different Fermi energies. This situation has
aroused great interest in the context of DC biased quantum
dots~\cite{KoG01}, where a  localized moment in the quantum dot is
coupled to two leads at different voltage bias. Recent research
have shown that magnetic moments, rather than electron-electron
interactions play the dominant role in the energy relaxation of
mesoscopic (``meso'') copper, gold and silver
wires~\cite{Pothier_exp,KG01,KZ02,GG01}. This  discovery  suggests
that considerable additional insight into the nonequilibrium
quantum physics of magnetic moments can be obtained from the study
of mesoscopic wires.

Evidence that magnetic moments are responsible for the energy
relaxation in mesowires comes from a variety of sources.  The
original experiments by the Saclay Group~\cite{Pothier_exp} showed
that the broadening of the two-step distribution function in a DC
biased mesowire involves a collision integral $K$ with energy
dependence $K\sim \epsilon^{-2}$. Kaminski and Glazman~\cite{KG01}
(KG) later identified this behavior as characteristic of spin-flip
scattering by magnetic impurities. Subsequent theoretical
work~\cite{KZ02,GG01}, and experimental confirmation~\cite{APPE03}
that magnetic fields quench the inelastic scattering~\cite{GGAG02}
have essentially confirmed  this hypothesis. One of the strong and
as yet untested consequences of this conclusion, is that at small
voltage biases, the spins inside a meso-wire should undergo the
``Kondo effect'', whereby the electron gas quenches the spin
replacing the strong inelastic scattering by phase-coherent
resonant scattering.

In this letter we consider the effects of reducing the voltage
bias across a meso-wire to a value comparable and then lower than
the characteristic Kondo temperature  of magnetic moments  in the
wire. We show that in the dilute limit the impurity physics in the
wire is isomorphic with that of a voltage biased quantum dot. The
low frequency shot noise measures the energy relaxation in the
wire, and it can be used as an alternative to the distribution
function in probing the inelastic processes in the wire. We
predict that the shot-noise to current ratio will peak as the 
bias $V$ is reduced
to the Kondo scale $T_K$, and will eventually decrease back to its
elastic limit as the bias is reduced even further. The predicted
noise peak reflects enhanced inelastic scattering at energies of 
the order of $T_K$\cite{Zarand04}.

Magnetic impurities are described by an Anderson model:
\begin{equation}
{\cal H}_{\rm imp}^i=  \sum_{\sigma}
  \epsilon_d n_i^{\sigma}+
Un_i^{\uparrow}n_i^{\downarrow} + W \sum_{\sigma} \left[
d^{\dagger}_{i \sigma} \psi_{\sigma}({\vec r}_i) +{\rm H.c.}
\right] ,
\end{equation}
where $d^{\dagger}_{i \sigma}$ creates a spin-$\sigma$ electron on
the $i$-th impurity,
$n_i^{\sigma}$ is the corresponding occupation number  and
$\psi^{\dagger}_{\sigma}(\vec r_i)$ creates a conduction electron
located at the same site.
The impurity parameters are
the orbital energy $\epsilon_d$, the on site repulsion $U$ and the
mixing amplitude $W$. Another important parameter is the
hybridization $\Gamma=\pi \rho W^2 \ll -\epsilon_d, \epsilon_d+U$,
where $\rho$ is the conduction band density of states. The
Hamiltonian of the wire is given by ${\cal H}={\cal H}_{\rm
CB}+{\cal H} _{\rm D}+\sum {\cal H}_{\rm imp}^i$, where the first
two terms describe the conduction band and the disorder potential
respectively.

The parameter which determines the amount of broadening in the
distribution function is $n_{\rm in}=\tau_D/ \tau_{\rm i}$, where
$\tau_D^{-1}=D/L^2$ is the Thouless energy and $\tau_{\rm i}^{-1}$
is the energy relaxation rate due to the magnetic impurities ($L$
and $D$ are the wire's length and diffusion coefficient). $n _{\rm
in}$ can be interpreted as the number of impurity mediated
inelastic scattering events that take place during a diffusive
pass across the sample. For $n_{\rm in} \ll 1$,  the wire is
mesoscopic and the distribution function is close to elastic
distribution, while for $n_{\rm in} \gg 1$ the wire is macroscopic
and the distribution function approaches a local equilibrium
distribution. We shall study the mesoscopic limit and calculate
the distribution and noise to first order in $n_{\rm in}$.

For a small source-drain bias $eV \ll k_BT_K$, the impurity is
screened and governed by Fermi-liquid physics. The relaxation rate
in this regime must increase quadratically with the bias $\tau
_{\rm i}^{-1} \propto V^2$. On the other hand, for the large bias
regime $eV \gg k_BT_K$, the Kondo singularities are cutoff by the
bias. As the bias is increased the Kondo correlations diminish and
as a result the effective coupling between the impurity and the
conduction electrons decreases logarithmically. The energy
relaxation rate in this regime should decrease like a polynomial
of the effective coupling $\ln^{-1} (eV/k_B T_K)$. Based on these
basic arguments the relaxation rate and $n _{\rm in}$ are maximal
at some intermediate bias of the order of $T_K$\cite{note1}. 

Suppose $x\in[0,1]$ is the fractional
distance along the wire from the left electrode.
In a wire without magnetic impurities the
distribution function is elastic: $f_x^{(0)}(\epsilon) =x
f_F(\epsilon - \mu_L)+(1-x)f_F(\epsilon-\mu_R)$, where
$\mu_{L(R)}=\mp eV/2$, and $f_F$ is the Fermi distribution. The
corresponding shot noise relates to the current $I$ through
$S^{(0)}=2eI/3$. In the dilute case, $n_{\rm in}\ll 1$, the
distribution and the noise can be expanded in the impurity
concentration $c_{\rm i}$:
\begin{eqnarray}
f&=&f^{(0)}+c_{\rm i}f^{(1)} + {\cal O}(c_{\rm i}^2), \nonumber \\
S&=&S^{(0)}+c_{\rm i} S^{(1)}+{\cal O}(c_{\rm i}^2).
\end{eqnarray}
The leading corrections in these expressions are the single impurity
contributions. The maximum in $n_{\rm in}$
will express itself in $S^{(1)}$ and in the amplitude of
$f_x^{(1)} (\epsilon)$.

The leading distribution correction is determined by making a
gradient expansion of the Dyson equations of the electron Greens
functions~\cite{RS86} which in turn yields the Boltzmann equation
\begin{equation}
-\frac{\partial ^2 f_x^{(1)}(\epsilon)}
{\partial x^2}=\tau_D I_{[f_x^{(0)}]} (\epsilon),
\label{Boltzmann}
\end{equation}
subject to boundary conditions $f^{(1)}|_{x=0,1}=0$. Here,
\begin{equation}
I_{[f_x^{(0)}]}(\epsilon) = -i \left\{ t_x^>(\epsilon)
f_x^{(0)} (\epsilon) + t_x^<(\epsilon) \left[
1-f_x^{(0)}(\epsilon) \right] \right\}, \label{Integral}
\end{equation} is the collision integral and
$t_x^{>(<)}$ is the greater (lesser) irreducible $t$-matrix of a
single impurity in a bath with an elastic distribution function
$f_x^{(0)}(\epsilon)$.

\begin{figure}
\centerline{
\includegraphics[width=80mm]{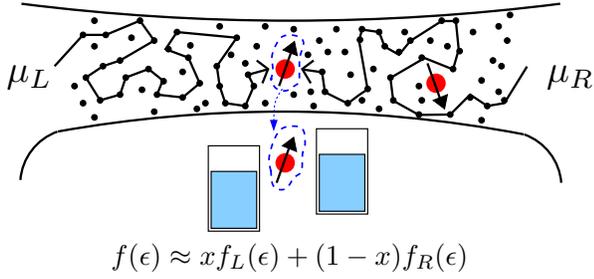}
}\vspace{-5pt} \caption{
        For the case of $n_{\rm in} \ll 1$, each electron is
        scattered inelastically at most by one magnetic impurity.
        As a result the diffusion due to the disorder potential
        (dots), communicates the energy distribution in the leads
        to the impurities, which act as if coupled to two distinct
        Fermi seas.}
\label{fig:fig1}
\end{figure}

The elastic distribution is composed of two Fermi distributions,
hence the $t$-matrices $t_x^<$ and $t_x^>$ are exactly equivalent
to $t$-matrices of a quantum dot which is coupled by tunneling
amplitudes $\sqrt{x} W$ and $\sqrt{1-x}W$ to two electrodes at
chemical potential $\mu_L$ and $\mu_R$ respectively. This proves
the equivalence of the wire problem to a quantum dot problem for
$n_{\rm in} \ll 1$. Figure \ref{fig:fig1} illustrates
schematically the relation between the problems.

We shall now apply a set of standard approximations to calculate
the $t$-matrices at the small and the large bias regimes and to
solve $f^{(1)}$. The noise correction $S^{(1)}$ will then be
calculated through the relation~\cite{Nagaev92}
\begin{equation}
S^{(1)}= \frac{4}{R} \int_0^1 dx \int d\epsilon \left[
1-2f^{(0)}_x(\epsilon) \right]  f_x^{(1)}(\epsilon),
\label{Nagaev}
\end{equation}
where $R$ is the resistance of the wire. Note that the non
equilibrium Kondo effect is still an open problem and up to date
there is no reliable description of the crossover between small
and large biases. We shall employ our results in the small and
large bias regimes to interpolate the physics in the 
crossover\cite{note1}.

{\em Small bias regime} - The small voltage and low temperatures
region may be treated by perturbating around the Fermi liquid
fixed point. To do this, we employ an extension of Hewson's
renormalized perturbation theory~\cite{RPT_refs} (RPT) to weak
departures from equilibrium~\cite{Oguri01}. The RPT describes the
vicinity of the strong coupling fixed point by an Anderson model
with renormalized parameters and counter terms to compensate for
the high-energy contributions to the self-energy and vertex parts.
For small departures from equilibrium we may use the same
renormalized Hamiltonian but with a non-equilibrium distribution
$f^{(0)}_x$ for the conduction bath.

\begin{figure}
\centerline{
\includegraphics[width=85mm]{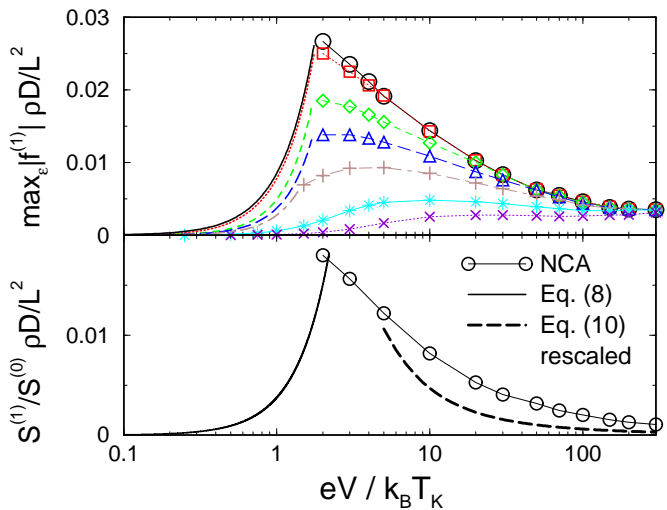}
}\vspace{-5pt}
\caption{
    Upper panel: amplitude of the distribution expansion in the middle
    of the wire $f^{(1)}_{x=1/2}$. The lines represent the small bias
    RPT expression of Eq.(\ref{RPT_f1}). The symbols with guide-lines
    show the large bias NCA results at temperatures of $T/T_K=0.001$,
    $0.01$, $0.05$, $0.1$, $0.2$, $0.5$ and $1$, displayed up to the
    intersection point with the RPT curves. Lower panel: shot-noise
    expansion $S^{(1)}$. The line stands for the small bias RPT
    expression of Eq.(\ref{RPT_dS}), the circles with guide-line
    describe the NCA results, and the thick dashed line shows the
    large bias asymptotic expression of Eq.(\ref{KG_dS}). The NCA
    parameters are ${\cal D}/ \Gamma=15$, ${\cal D}/\epsilon_d=-3.593$,
    where $2{\cal D}$ is the bandwidth~\cite{note2}.}
\label{fig:fig2}
\end{figure}

To leading order in $\epsilon /k_BT_K$, $eV / k_B T_K$ and $T /
T_K$, the generalized RPT calculation yields a self-energy with
the same functional dependence on energy and voltage as a
perturbative treatment of direct electron-electron scattering
\cite{Nagaev95}. In particular, the energy relaxation rate is
\begin{equation}
\frac{1}{\tau_{\rm i}^{\rm RPT}} = \frac{\pi c_{\rm i}}{16 \rho}
\left( \frac{eV}{k_BT_K} \right)^2.
\end{equation}

\begin{figure}
\centerline{
\includegraphics[width=85mm]{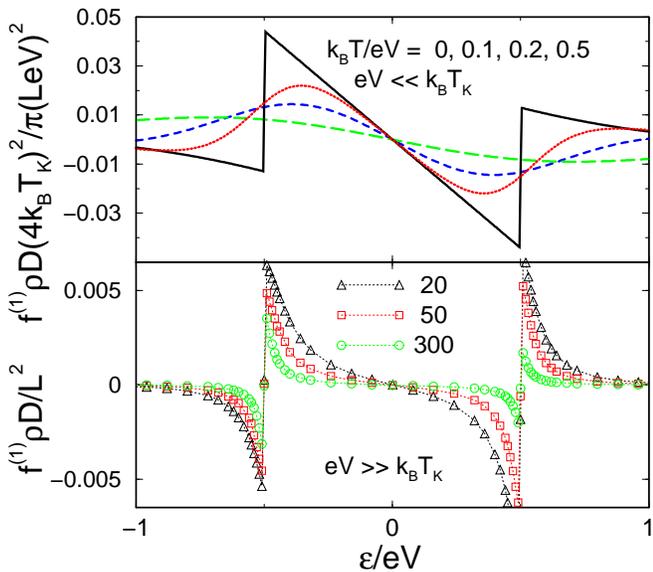}
}\vspace{-5pt}
\caption{
    The distribution expansion in the middle of the wire $f^{(1)}_{x=
    1/2}$. Upper panel: The small bias RPT expression of Eq.(\ref{RPT_f1})
    at different temperatures. For a given $T/V$ ratio the amplitude
    of $f^{(1)}$ is proportional to $(V/T_K)^2$ and to $L^2 /\rho D$.
    Lower panel: NCA results at $eV/k_B T_K= 20$, $50$ and $300$ for
    $T/T_K=0.001$. The parameters are the same as in Fig.~\ref{fig:fig2}. }
\label{fig:fig3}
\end{figure}

The solution of Eq.(\ref{Boltzmann}) with the collision integral
calculated by the RPT gives the distribution correction to second
order in $\epsilon /k_BT_K$, $eV/k_BT_K$ and $T/T_K$:
\begin{eqnarray}
f_x^{(1)}(\epsilon)= -\frac{\pi L^2}{16 \rho D}
\sum_{\ell=-3/2}^{3/2} q_{\ell}(x) \left[ \left( \frac{\pi T}{T_K}
\right)^2+\left( \frac{\epsilon -\ell eV}{k_BT_K} \right)^2 \right]
\nonumber \\
\times \left[ f^{(0)}_x(\epsilon) - f_F(\epsilon-\ell eV) \right].
\ \ \ \label{RPT_f1}
\end{eqnarray}
The spatial dependence enters through the polynomials
$q_{\frac{3}{2}}(x)=\frac{1}{20}(x-x^5) +\frac{1}{6}(x^4-x^3)$,
$q_{\frac{1}{2}}(x)=\frac{3}{20}x^5 - \frac{5}{12}x^4 +
\frac{1}{2} (x^3-x^2) + \frac{4}{15} x $, and their symmetric
counterparts $q_{-\ell} (x) = q_{\ell} (1-x) $. The contribution
of the magnetic impurities to the noise is
\begin{equation}
\frac{c_{\rm i}S^{(1)}}{S^{(0)}} = \frac{c_{\rm i} L^2}{\rho D}
\frac{19}{45\cdot 7} \frac{ \pi  }
{16 } \left(\frac{eV}{k_B T_K}\right)^2.
\label{RPT_dS}
\end{equation}

{\em Large bias regime} - We treat the large bias regime both
numerically, using the NCA, and analytically employing a scaling
argument proposed by KG. Numerically, it is convenient to
calculate the impurity spectral function $A_d$ and its effective
distribution $F_d=G^<_d/(2\pi i A_d)$, where $G_d^<$ is the
impurity lesser Green function.  The impurity is considered within
the NCA as an infinite-$U$ Anderson impurity, coupled to a
conduction bath with a non equilibrium distribution $f_x^{(0)}$.
In terms of these quantities the collision integral is given by $2
\rho^{-1} \Gamma A_d ( F_d - f^{(0)})$. The NCA takes account of
the inelastic scattering to produce a non-trivial impurity
distribution function $F_d \ne f^{(0)}$ (in contrast e.g. to
slave-boson mean-field) and provides a reliable description of the
large bias regime. However, the NCA fails to reproduce the
Fermi-liquid and cannot be fully extended to the small-bias,
low-temperature regime.

Analytically, following KG, we turn now to the Kondo model
rescaling the leading order perturbation theory in the exchange
coupling expression by replacing $\rho J \rightarrow \ln
^{-1}(eV/k_BT_K)$. This rescaling procedure describes the large
bias asymptotic behavior at $eV \gg k_BT_K$. The leading amplitude
of a spin mediated electron-electron inelastic process is of
second order in $\rho J$. For the sake of simplicity the spin
Green function is uniformly broadened with the Korringa rate
$\tau_S^{-1} = \zeta V (\rho J)^2$, where $\zeta$ is a factor of
order unity. Generally, $\zeta$ has a spatial
dependence~\cite{KZ02}, however the correct physics is still
captured by taking its value in the middle of the wire $\zeta=\pi
/4$. To leading order in $\rho J$ the collision integral kernel
reads:
\begin{equation}
K(\omega)=\frac{\pi }{2 \rho}  \frac{3}{4}
\frac{(\rho J)^4}{\omega^2+\tau_S^{-2}},
\end{equation}
where $\omega$ is the energy transferred between the interacting
electrons. The corresponding noise correction reads
\begin{eqnarray}
\frac{c_{\rm i} S^{(1)}}{S^{(0)}} = \frac{L^2 c_{\rm i}}{\rho D}
\frac{3\pi \eta^2} {80 \zeta^2}\left[ {\rm Re} \left\{
\left(1-i\eta \right) \ln \frac{\eta+i}{\eta} \right\}
-\frac{22}{21}\right], \label{KG_dS}
\end{eqnarray}
where $\eta \equiv \zeta (\rho J)^2$. Rescaling this expression we
find that the impurity contribution to the noise decays at large
biases like $\ln^{-4} (V/T_K)$.

Figure  \ref{fig:fig2} shows the amplitude of the first order
expansion of the distribution $f^{(1)}$ and the first order
expansion of the shot noise $S^{(1)}$. Both quantities show a peak
at intermediate biases of the order of the Kondo temperature
representing the enhanced inelastic scattering rate in the
crossover. The intensity of the relative contribution of the
magnetic impurities to both quantities is proportional to $c_{\rm
i}L^2/\rho D$. The amplitude of the distribution correction has a
strong temperature dependence and the peak is almost washed out at
temperatures of the order of $T_K$. The RPT and NCA results
intersect in the cross-over regime where the voltage is comparable
with the Kondo temperature.

Figure \ref{fig:fig3} shows the correction of the distribution
function in the middle of the wire $f^{(1)}_{x=1/2}$. This
quantity measures the smearing of  the two-step energy
distribution by inelastic impurity scattering. The correction
$f^{(1)}_{1/2}$ evolves with voltage and becomes sharper, gaining
an additional scale- $\eta \propto \ln^{-2} (eV/k_BT_K)$ at the
large bias regime.

\begin{figure}
\centerline{ 
\includegraphics[width=75mm]{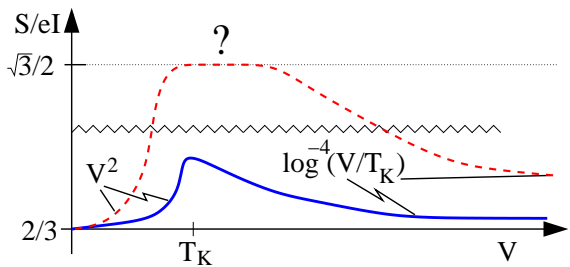}
}\vspace{-5pt} \caption{
    The perturbative arguments hold as long as $c_{\rm i}L^2/\rho D$
    is small, where the curve-shape of the noise is represented
    schematically by the solid curve. When the impurity concentration
    or wire's length are increased the perturbative description
    breaks down at intermediate biases. Assuming that the noise has a
    monotonic dependence on the inelastic scattering rate, we speculate
    that the noise will crossover to its $\sqrt{3}eI/2$ value of local
    equilibrium at the intermediate biases as illustrated by the dashed
    curve.}
\label{fig:fig0}
\end{figure}

According to Eqs.(\ref{Boltzmann}) and (\ref{Nagaev}) the small
parameter of the $c_{\rm i}$ expansion is $\lambda = c_{\rm i}
\tau_D \max_{\epsilon} I_{[f^{(0)}]}$. $\lambda$ acquires its
maximum value at the noise peak, so we can estimate its upper
bound from the  RPT value at $eV=k_BT_K$:
\begin{eqnarray}
\lambda(V) \le \lambda^{\rm RPT} (k_BT_K/e) = \frac{c_{\rm i}
L^2}{\rho D} \frac{6\pi}{256} \approx 0.07 \frac{c_{\rm i}
L^2}{\rho D}. \label{bound}
\end{eqnarray}
For wires in which this bound is smaller than $1$, the
distribution and the noise are well described by perturbation
theory. The noise curve of such a wire is represented
schematically by the solid curve of Fig.\ref{fig:fig0}. For wires
in which $c_{\rm i} L^2/\rho D \gtrsim 14$ the perturbative
approach breaks down at intermediate biases. At very small or very
large biases the energy relaxation is still small enough hence the
curve will preserve its $V^2$ or $\ln^{-4}(eV/k_BT_K)$
asymptotics. At intermediate-small or -large biases perturbation 
theory breaks down in this case and one needs to solve the Boltzmann 
equation  self-consistently like in Refs.\cite{KZ02,Ujsaghy04}.
We can however speculate that in this bias regimes the wire is well
described by local equilibrium and that the noise crosses over
from its local equilibrium value of $\sqrt{3}eI/2$ as presented
schematically by the dashed curve in Fig.\ref{fig:fig0}.

At present we lack a precise theoretical description of the nonequilibrium
crossover regime and the detailed shape of the noise peak in this regime. In
particular, it is not yet clear whether the cross-over will involve a broad
maximum in the noise, reminiscent of the energy dependent inelastic scattering
rate in the ground-state of the Kondo model\cite{Zarand04}, or a sharper 
maximum, as observed experimentally in the temperature dependent dephasing 
rate\cite{dephasing_exp}. This issue awaits future theoretical and 
experimental resolution.

An interesting question for the future concerns the extreme case
of a heavy electron meso-wire,  made up of a dense array of local
moments which are quenched at zero voltage bias. Would it be
possible to realize a non- equilibrium distribution in such a
wire, and if so how would the non equilibrium state effect the
competition between RKKY- mediated magnetism and Kondo lattice
formation~\cite{VGL03}? This is not an abstract question, for
micron size filaments of  the heavy electron ${\rm UPt}_3$ are
formed naturally in the melt\cite{Taillefer87}.

This paper has examined the shot-noise in diffusive wires
containing dilute impurities, showing that in the dilute limit the
behavior of magnetic impurities inside the meso-wire is equivalent
to the physics of DC biased quantum dot. The low frequency noise
appears to be an ideal probe of the inelastic processes in such a
non equilibrium Kondo system. We have calculated the distribution
function and noise at the small and the large bias regimes. A key 
prediction of this paper is the appearance of a noise peak in the 
nonequilibrium Kondo crossover at voltages $V \sim T_K$. 
While many other sources of noise may exhibit a voltage
dependence, this important contribution to the noise will 
uniquely depend on the concentration of impurities, and will also 
exhibit a magnetic field dependence.

Acknowledgments: We thank D. Langreth, A. Schiller, A. Silva, and
M. M\"uller, for discussions. This work is supported by DOE grant
DE-FE02-00ER45790.



\begin{thebibliography}{99}
\bibitem{note1}
    {The only scale is the Kondo temperature hence no additional 
    structure to the maximum peak is expected.}
\bibitem{note2}
    {The NCA results are almost unchanged
    when choosing different model parameters within the
    Kondo regime.}
\end{thebibliography}
\end{document}